\documentclass[a4paper]{article}
\usepackage{spconf}

\usepackage{amsmath}
\usepackage{hyperref}
\usepackage{enumitem}
\usepackage{graphicx}
\usepackage{grffile}
\usepackage{multirow}
\usepackage{multicol}
\usepackage{xcolor}

\usepackage[font=scriptsize]{subcaption}

\newcommand{\tacotron}{\textsc{s2s}}
\newcommand{\autoregressive}{{\footnotesize \textsc{DurIAN+}}}
\newcommand{\ptm}{\textsc{camp}}
\newcommand{\oracle}{\textsc{ora}}

\newcommand{\speech}{\textsc{nat}}

\definecolor{PhoneGreen}{HTML}{4A8547}
\definecolor{DurationYellow}{HTML}{CC9900}
\definecolor{AcousticBlue}{HTML}{0F58BD}
\definecolor{ContextPurple}{HTML}{6F4A82}
\definecolor{ProsodyOrange}{HTML}{CC6900}

\makeatletter
\def\blfootnote{\xdef\@thefnmark{}\@footnotetext}
\makeatother

\let\OLDthebibliography\thebibliography
\renewcommand\thebibliography[1]{
  \OLDthebibliography{#1}
  \setlength{\parskip}{0.5pt}
  \setlength{\itemsep}{0.5pt plus 0.3ex}
}

\title{CAMP: A two-stage approach to modelling prosody in context}

\name{Zack Hodari\sthanks{Work performed during an internship at Amazon TTS Research.}, Alexis Moinet, Sri Karlapati, Jaime Lorenzo-Trueba, Thomas Merritt, Arnaud Joly,}
\nameplus{Ammar Abbas, Penny Karanasou, Thomas Drugman\vspace{-5pt}}

\address{
  Amazon Research, Cambridge, United Kingdom\\
  *The Centre for Speech Technology Research, University of Edinburgh, United Kingdom\\
	\fontsize{9}{9}\selectfont\ttfamily\upshape
	zack.hodari@ed.ac.uk, \{amoinet, srikarla, drugman\}@amazon.com
	\vspace{-10pt}
}

\begin{document}

\maketitle

\begin{abstract}
\vspace{-4pt}
Prosody is an integral part of communication, but remains an open problem in state-of-the-art speech synthesis. There are two major issues faced when modelling prosody: (1) prosody varies at a slower rate compared with other content in the acoustic signal (e.g.\ segmental information and background noise); (2) determining appropriate prosody without sufficient context is an ill-posed problem. In this paper, we propose solutions to both these issues. To mitigate the challenge of modelling a slow-varying signal, we learn to disentangle prosodic information using a word level representation. To alleviate the ill-posed nature of prosody modelling, we use syntactic and semantic information derived from text to learn a context-dependent prior over our prosodic space. Our context-aware model of prosody (CAMP) outperforms the state-of-the-art technique, closing the gap with natural speech by 26\%. We also find that replacing attention with a jointly-trained duration model improves prosody significantly.
\end{abstract}


\vspace{-6pt}
\section{Introduction}
\label{sec:intro}
\vspace{-10pt}

Producing realistic speech with text-to-speech (TTS) synthesis can be split into two goals: naturalness and appropriateness. We define naturalness as similarity of acoustic quality to human speech, and appropriateness as a measure of how well a prosodic rendition fits the given context -- where context refers to any information used in prosody planning.

While the naturalness of state-of-the-art TTS is near identical to human speech \cite{transformer-tts:2019}, the prosody may not always be realistic \cite{srikanth-thesis:2019}. Overall, prosody in TTS might be described as boring or flat \cite{zack-SSW19:2019}, and can become fatiguing for listeners \cite{avashna:2018}.

Unrealistic prosody modelling is often linked to a lack of contextual information. Indeed, without sufficient context, predicting prosody is an ill-posed problem \cite{rob-SSW19:2019}, as any number of prosodies could be deemed appropriate for a given text. We believe that this is the biggest limitation of current state-of-the-art TTS models. While it may be possible to learn prosody implicitly, generating appropriate prosody without additional context information is challenging. In this paper, we consider suprasegmental prosody: prosodic phenomena above the phonetic level \cite{ladd:2008,ward:2019}. Suprasegmental prosody is influenced by a broad range of context information, from syntax and semantics to affect, pragmatics, and setting \cite{intonational-bestiary:2016,syntax-prosody-correlation:2018}.

Modelling suprasegmental prosody is made more difficult due to its medium; speech carries segmental information, background noise, and prosody in parallel \cite{ladd:2008}. Suprasegmental prosody exists at a different resolution to these as it varies more slowly \cite{ward:2019}, making it challenging to model explicitly.

To deconstruct these problems, we propose a novel two-stage approach: context-aware modelling of prosody (CAMP).
In \emph{stage-1}, a prosody representation is learnt, and in \emph{stage-2}, that representation is predicted using context. We also improve Tacotron using a jointly-trained duration model.

Representation learning for speech has been investigated using unsupervised methods \cite{VQ-VAE:2017,wav2vec:2019,zerospeech:2019}. While there are many methods to define prosodic correlates \cite{suni-representations:2015,aggregated-prosody:2019,sam-wavelets-DCT:2015}, there is less work on unsupervised representation learning for prosody specifically. Most prosody representations are learnt at the sentence-level \cite{tacotron-GST:2018,amazon-dynamic-prosody:2020}. However, these are too coarse and are not able to perfectly reconstruct prosody \cite{xin-F0-VQ-VAE:2019}. To accurately capture prosody we need a sequence of representations, e.g.\ word or phrase level \cite{xin-F0-VQ-VAE:2019,zack-SP20:2020}. Following CopyCat \cite{copycat:2020}, we learn a sequence of representations from the mel-spectrogram, but we do so at the word-level (\emph{stage-1}).

As demonstrated by VQ-VAE \cite{VQ-VAE:2017}, a learnt prior is capable of unsupervised phonetic discovery when using a high bit rate representation. Reducing the bit rate of the representation, by using a longer unit length, should allow the learnt prior to capture longer-range effects, such as prosodic patterns. The linguistic linker introduced by Wang et al.\ \cite{xin-F0-VQ-VAE:2019} experiments with different unit lengths when modelling F\textsubscript{0}, but not other aspects of prosody such as rhythm and intensity. Representations extracted from a spectrogram can capture these aspects of prosody \cite{tacotron-TP-GST:2018,amazon-dynamic-prosody:2020}. TP-GST \cite{tacotron-TP-GST:2018} predicts embeddings, but only using segmental information. Tyagi et al.\ \cite{amazon-dynamic-prosody:2020} use additional context information, but don't train a prediction model. We argue that additional context information is vital to improving prosody quality in TTS. As such, we propose \emph{stage-2}: use suprasegmental context information to predict the sequence of prosody representations from \emph{stage-1}.

In Section~\ref{sec:AR-TTS}, we present our baselines. In Section~\ref{sec:method}, we describe the two-stage training of CAMP.

\section{Baseline sequence-to-sequence TTS}
\label{sec:AR-TTS}
\vspace{-5pt}

\textbf{Attention-based model --- }We use Tacotron-2 \cite{tacotron-2:2018} as our attention-based baseline, \tacotron. Tacotron-2 consists of a phoneme encoder, attended over by an auto-regressive acoustic decoder using location-sensitive attention.

\noindent\textbf{Duration-based model --- }Attention serves two purposes in TTS: prediction of rhythm and summarisation of relevant phonetic context. FastSpeech-2 \cite{fastspeech-2:2020} and DurIAN \cite{DurIAN:2019} demonstrated that: a duration prediction model can handle rhythm; and convolutions, self-attention, or bi-directional recurrent layers can summarise local context. We introduce a second baseline, \autoregressive, that uses a phone duration model instead of attention. \autoregressive\ is similar to DurIAN but, like FastSpeech-2, our duration model is trained jointly and shares a phone encoder with the acoustic decoder. Thus, the phone embeddings are influenced by both acoustic and duration losses, in turn providing richer inputs to the duration model at inference time. Unlike DurIAN, \autoregressive\ does not use style control or prosodic boundary tokens.\footnote{The comparison with DurIAN is intended to link \autoregressive\ to duration-based models in the literature. In reality, \autoregressive\ is very similar to \tacotron.} As shown in Section~\ref{sec:evaluations-AR}, it is a competitive state-of-the-art baseline.

\begin{figure}[t]
  \centering
  \includegraphics[width=1.15\linewidth]{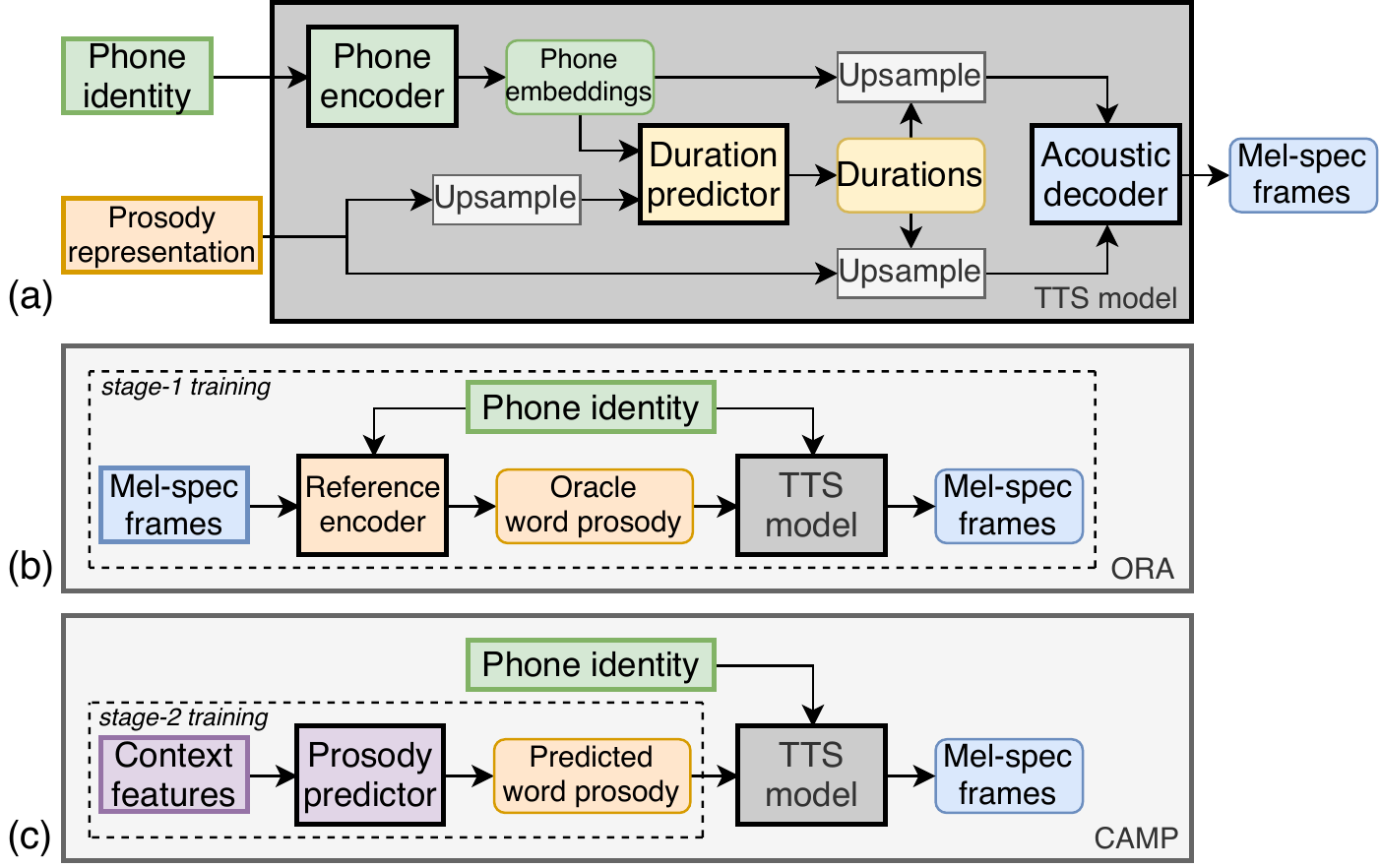}
  \caption{(a) TTS model -- synthesis driven by a prosody representation. (b) \oracle\ -- uses prosody copied from speech. (c) \ptm\ -- uses prosody predicted from context. Dashed boxes in (b) and (c) correspond to two training stages in Section~\ref{sec:method}.}
  \label{fig:TTS}
  \vspace{-10pt}
\end{figure}

\begin{figure}[t]
  \hspace{17pt}
  \begin{subfigure}[b]{0.49\linewidth}
    \centering
    \includegraphics[width=\linewidth]{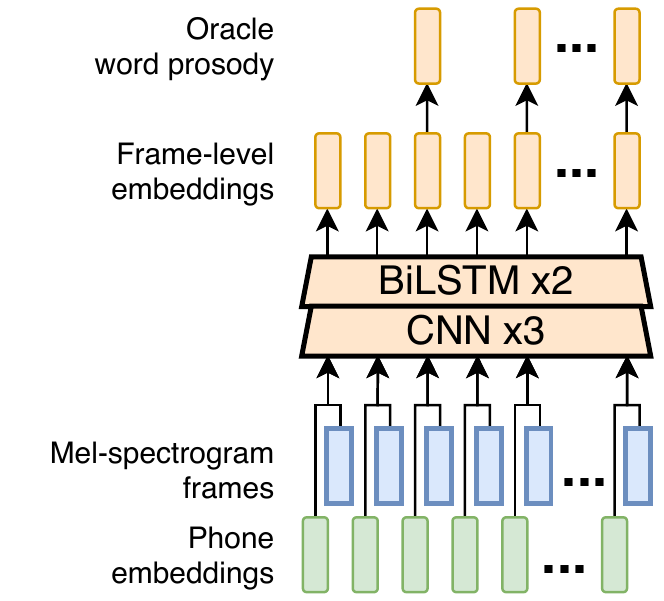}
    \caption{\emph{Word-level} reference encoder}
    \label{fig:modules-reference}
  \end{subfigure}
  \begin{subfigure}[b]{0.42\linewidth}
  \hspace{-10pt}
    \centering
    \includegraphics[width=\linewidth]{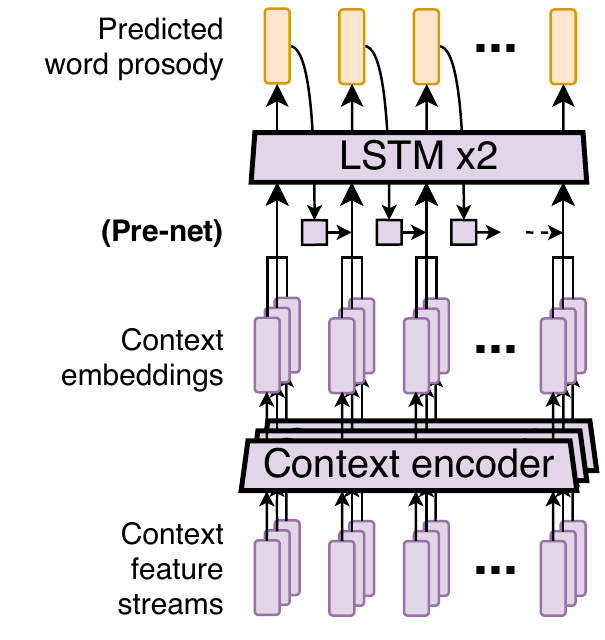}
    \caption{Prosody predictor}
    \label{fig:modules-prosody}
  \end{subfigure}
  \begin{subfigure}[b]{0.43\linewidth}
    \centering
    \includegraphics[width=\linewidth]{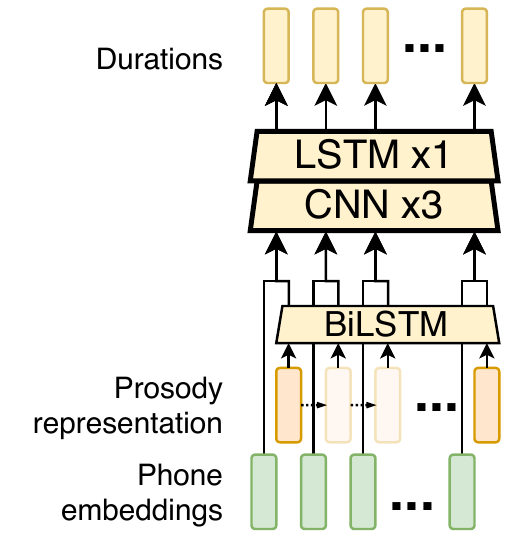}
    \caption{Duration predictor}
    \label{fig:modules-duration}
  \end{subfigure}
  \begin{subfigure}[b]{0.54\linewidth}
    \centering
    \includegraphics[width=\linewidth]{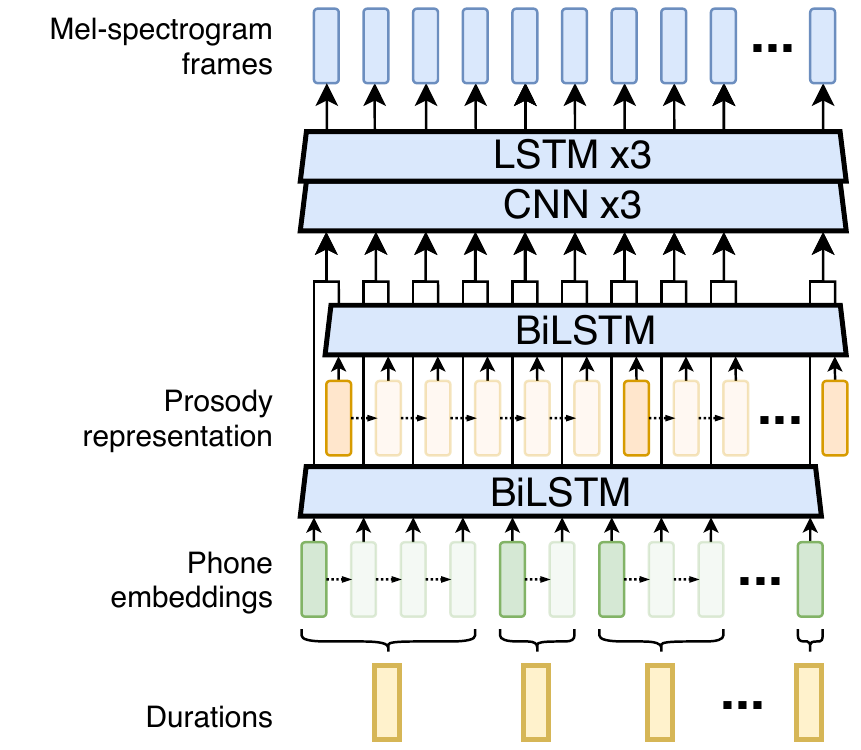}
    \caption{Acoustic decoder}
    \label{fig:modules-acoustic}
  \end{subfigure}
  \caption{Details of modules in Figure~\ref{fig:TTS}. (a) Represents prosody at the word level. (b) Auto-regressively predicts prosody representations using context features. (c) \& (d) predict their targets using either oracle or predicted prosody representations.}
  \label{fig:modules}
  \vspace{-12pt}
\end{figure}

\vspace{-4pt}
\section{Two-stage prosody modelling}
\label{sec:method}
\vspace{-6pt}

Our proposed approach, \ptm\ (Figure~\ref{fig:TTS}c), predicts the prosody representation used to drive a \emph{``TTS model''}, detailed in Figure~\ref{fig:TTS}a. In section~\ref{sec:repr} we discuss \emph{stage-1} of training, where this \emph{TTS model} is trained using a word-level reference encoder, with the goal of perfectly reconstructing the original prosody. \oracle, the oracle TTS, is the \emph{TTS model} when driven by this reference encoder (Figure~\ref{fig:TTS}b). Section~\ref{sec:prosody} describes \emph{stage-2}, where we train a prosody predictor that aims to predict the oracle prosody representations, using suprasegmental text-derived context features. Training these stages separately ensures that the context is used to predict prosody, not lower-level acoustic information. It is essential to context-aware prosody modelling that context information feeds into a loss that explicitly focusses on prosody.

\vspace{-5pt}
\subsection{Word-level prosodic representation learning}
\label{sec:repr}
\vspace{-5pt}

To learn a disentangled representation of suprasegmental prosody, mel-spectrograms are auto-encoded with a temporal bottleneck \cite{xin-F0-VQ-VAE:2019,copycat:2020} (Figure~\ref{fig:TTS}b). The temporal bottleneck limits the model to only select information not predictable from other inputs, i.e.\ phone identity. However, determining the unit length for the bottleneck involves a trade-off: too short a unit length and segmental and background information will remain entangled \cite{VQ-VAE:2017}; too long and we will lose descriptive power \cite{tacotron-GST:2018,CHiVE:2019}. Since sentences contain an arbitrary amount of information \cite{zack-SP20:2020}, we use words as the unit length in the temporal bottleneck. Specifically, we use the final embedding for each word and each pause (Figure~\ref{fig:modules-reference}) as the oracle representation to be learned by \emph{stage-2}.

The duration model trained in \emph{stage-1} (Figure-\ref{fig:modules-duration}) uses phone embeddings \emph{and} our disentangled prosody representation (upsampled to phone-level) as input. This ensures our representation can capture duration-related information.

The acoustic decoder (Figure~\ref{fig:modules-acoustic}) is similar to CopyCat's decoder \cite{copycat:2020}, but with BiLSTMs to smooth upsampled features. It uses no auto-regressive feedback of predictions, unlike the decoders used by the models discussed in Section~\ref{sec:AR-TTS}.

\subsection{Context-aware prosody prediction}
\label{sec:prosody}
\vspace{-2pt}

We train a prosody predictor (Figures~\ref{fig:TTS}c and \ref{fig:modules-prosody}) to predict the word-level prosody representations described in Section~\ref{sec:repr}, using text-derived features. This task has been referred to as ``linguistic linking'' \cite{xin-F0-VQ-VAE:2019}, or ``text-predicted tokens'' \cite{tacotron-TP-GST:2018}. We build upon this idea, emphasising the need for suprasegmental features to see improvement on this task.

To take advantage of the correlation between syntax and prosody \cite{syntax-prosody-correlation:2018}, we experiment with four syntax-related context features. Each syntax feature uses a separate \emph{context encoder}, consisting of 2 CNN layers, followed by a BiLSTM.
\vspace{-4pt}
\begin{itemize}[itemsep=-4pt,leftmargin=10pt]
  \item \textbf{Part-of-speech} (POS) -- syntactic role of a word. We use the LAPOS tagger \cite{LAPOS:2011} with the Penn-Treebank tagset.
  \item \textbf{Word-class} -- open or closed class (content or function).
  \item \textbf{Compound-noun structure} -- flag indicating if a word is part of a compound noun. Extracted using POS tags.
  \item \textbf{Punctuation structure} -- flag indicating punctuation.
\end{itemize}
\vspace{-2pt}

We use one semantic \emph{context encoder}: a pre-trained BERT\textsubscript{BASE} model \cite{BERT:2019}. During \emph{stage-2} of training this is fine-tuned on our prosody prediction task. BERT's word-piece embeddings are aligned to word-level by average pooling.

Our prosody predictor is auto-regressive (illustrated by the curved downwards arrows in Figure~\ref{fig:modules-prosody}). This allows the model to learn how prosody transitions from one word to the next. Adding attention or self-attention might improve modelling of distant information. However, we believe this is less important as our sentences are only on average 16 words.

Our prosody predictor can be interpreted as a learnt conditional prior, where the latent space -- our prosody representation -- has unknown variance. Unlike VQ-VAE \cite{VQ-VAE:2017}, we use a coarser-grained representation (i.e.\ word-level), allowing us to focus on prosody, as opposed to frame-level information.

\vspace{-5pt}
\section{Experiments}
\vspace{-5pt}

\subsection{Data}
\label{sec:data}
\vspace{-5pt}

We train our models on prosodically rich data: an expressive single-speaker internal dataset. The data contains about 25,000 utterances of professionally recorded speech by a native US English female speaker. The training, validation, and test sets are approximately 30 hours, 2 hours, and 6 hours, respectively. Phoneme features are one-hot encodings of phone identity and service tokens, such as word-boundaries. We extracted 80-band mel-spectrograms with a 12.5 ms frame-shift. Durations were extracted using forced alignment with Kaldi.

\vspace{-8pt}
\subsection{Systems}
\label{sec:systems}
\vspace{-5pt}

We use the two single-stage baselines described in Section~\ref{sec:AR-TTS}: \tacotron\ (Tacotron-2 \cite{tacotron-2:2018}); and \autoregressive\ (\tacotron\ with a jointly-trained duration model). Table~\ref{tab:systems} details \autoregressive, along with our two-stage models; \ptm, and \oracle. Our proposed model, \ptm, uses prosody predicted from context features (Figure~\ref{fig:TTS}c). \oracle\ represents the top-line performance of our two-stage approach -- using oracle prosody copied from human speech (Figure~\ref{fig:TTS}b). All models use the same 24kHz mixture of logistics WaveNet vocoder \cite{parallel-wavenet:2018}. \speech\ is natural 24kHz speech with no vocoding, used in Section~\ref{sec:evaluations-CAMP}.

All models were trained using the Adam optimiser. An L1 loss was used for acoustic and duration losses. \tacotron, \autoregressive, and \oracle\ were trained for 300,000 steps, with a learning rate of 0.001 and a decay factor of 0.98. Prosody prediction used a Huber loss with $\rho = 1$, this produced marginally better prosody than an L1 or L2 loss during informal listening. Prosody predictors were trained for 100,000 steps, with a learning rate of 0.0001 and a decay factor of 0.98.

\begin{table}[b]
\begin{center}
  \renewcommand{\arraystretch}{1.1}
  \vspace{-8pt}
  \caption{System comparison. \autoregressive\ is a competitive baseline. \oracle\ represents our approaches' best-case performance. \ptm\ is our proposed model using context information.}
  \vspace{-4pt}
  \begin{tabular}{ |l|ccc| } \hline
                                   & \autoregressive      & \oracle                            & \ptm \\ \hline\hline
    \textcolor{PhoneGreen}{Phone encoder} & \multicolumn{3}{c|}{Tacotron-2 encoder}                    \\ \hline
    \multirow{2}{*}{\textcolor{ProsodyOrange}{Prosody}}       & \multirow{3}{*}{---} & \textcolor{ProsodyOrange}{Reference} & \textcolor{ContextPurple}{Prosody} \\
    \multirow{2}{*}{\textcolor{ProsodyOrange}{representation}} &                     & \textcolor{ProsodyOrange}{encoder} & \textcolor{ContextPurple}{predictor} \\ 
                                   &                      & (Fig.~\ref{fig:modules-reference}) & (Fig.~\ref{fig:modules-prosody})     \\ \hline
    \textcolor{DurationYellow}{Duration} & \textcolor{PhoneGreen}{Phone}      & \multicolumn{2}{c|}{\textcolor{PhoneGreen}{Phone embeddings} and}     \\
    \textcolor{DurationYellow}{inputs}   & \textcolor{PhoneGreen}{embeddings} & \multicolumn{2}{c|}{\textcolor{ProsodyOrange}{prosody representation}} \\ \hline
    \textcolor{AcousticBlue}{Acoustic} & Tacotron-2 & \multicolumn{2}{c|}{\textcolor{AcousticBlue}{CopyCat decoder}} \\
    \textcolor{AcousticBlue}{Decoder}  & decoder    & \multicolumn{2}{c|}{(Fig.~\ref{fig:modules-acoustic})} \\ \hline
  \end{tabular}
  \label{tab:systems}
\end{center}
\vspace{-20pt}
\end{table}

\vspace{-8pt}
\subsection{Subjective evaluation}
\label{sec:evaluations}
\vspace{-4pt}

For our listening tests we use 96 sentences, chosen randomly from the test set. In Section~\ref{sec:evaluations-ablation}, we evaluate the efficacy of our syntactic and semantic features. Following this, we measure the contribution of duration modelling (Section~\ref{sec:evaluations-AR}). This allows us to control for the effect of duration modelling when evaluating our core contribution in Section~\ref{sec:evaluations-CAMP}.

\vspace{-8pt}
\subsubsection{Context features}
\label{sec:evaluations-ablation}
\vspace{-5pt}

In Section~\ref{sec:prosody}, we described 5 different context features. To determine the contribution of each to prosody performance we conduct an ablation test using a MUSHRA-like format (no anchor and no reference) \cite{MUSHRA:2014}. We consider three separately trained models: \ptm\textsubscript{Syntax} -- using 4 syntax context encoders, \ptm\textsubscript{BERT} -- using the BERT context encoder, and \ptm\textsubscript{BERT+Syntax} -- using all 5 context encoders. 20 listeners compared these three \ptm\ models on a scale from 0 to 100.

We perform a two-sided Wilcoxon signed-rank test on all 3 pairs of systems, and correct with Holm-Bonferroni. The results in Table~\ref{tab:ablation} show that \ptm\textsubscript{Syntax} is significantly worse than both other systems ($p < 0.001$). No statistically significant difference is found between \ptm\textsubscript{BERT} and \ptm\textsubscript{BERT+Syntax} ($p = 0.83$). We conclude that our syntactic features provide no additional information compared to BERT's contextualised representations. This is in agreement with findings that BERT can capture both semantic \emph{and} syntactic information \cite{BERTology:2020}. As such we use \ptm\textsubscript{BERT} as our proposed system (\ptm\ for brevity).

\begin{table}[t]
  \begin{center}
  \renewcommand{\arraystretch}{1.1}
	  \caption{MUSHRA ablation of context features in \ptm. Mean rating and 95\% confidence intervals are reported. Question asked: \emph{``Rate the systems based on your preference''}.}
  \begin{tabular}{ |r|lll| } \hline
                 & Syntax & BERT & BERT+syntax \\
    Mean score   & 63.9 $\pm$1.21 & 66.5 $\pm$1.21 & 66.3 $\pm$1.19 \\ \hline
  \end{tabular}
  \label{tab:ablation}
  \end{center}
  \vspace{-10pt}
\end{table}

\subsubsection{Benchmark model}
\label{sec:evaluations-AR}
\vspace{-5pt}

We compare the attention-based \tacotron\ with the duration-based \autoregressive\ to measure what improvement our duration model provides. We perform a preference test with 15 listeners.

\begin{figure}[t]
  \centering
  \includegraphics[width=\linewidth]{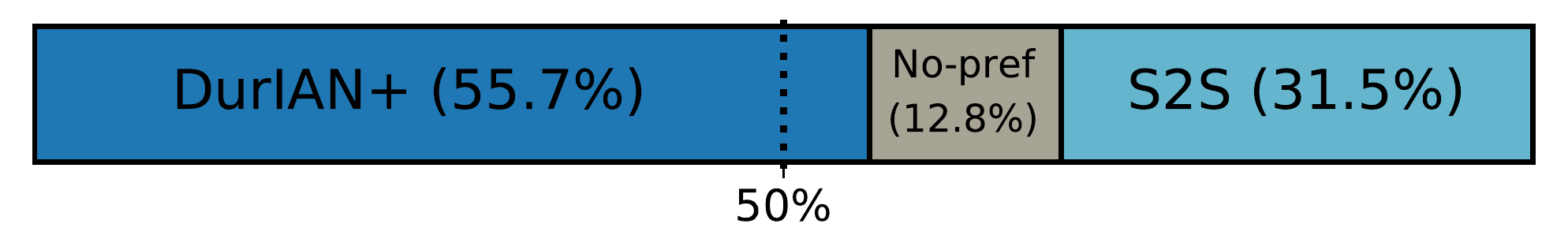}
  \vspace{-20pt}
	\caption{Preference test demonstrating that a jointly-trained duration model improves prosody significantly over Tacotron-2. Question asked: \emph{``Choose which version you prefer''}.}
  \label{fig:preference}
  \vspace{-10pt}
\end{figure}

To show that \autoregressive\ improves prosody, not just robustness, we remove the 20 sentences where \tacotron\ had attention instabilities from the results presented in Figure~\ref{fig:preference}. There were no sentences where \autoregressive\ had stability issues. A binomial significance test demonstrates that \autoregressive\ is significantly preferred over \tacotron\ ($p < 10^{-15}$).

We tried training \autoregressive\ with a fine-tuned BERT as an additional encoder. However, there was little or no improvement in prosody -- similar to previous findings \cite{tomoki-BERT-TTS:2019,glass-BERT-TTS:2019}. We believe this is due to this single-stage model predicting low-level acoustics, as opposed to a prosodic representation.

\begin{figure}[t]
  \centering
  \includegraphics[width=\linewidth]{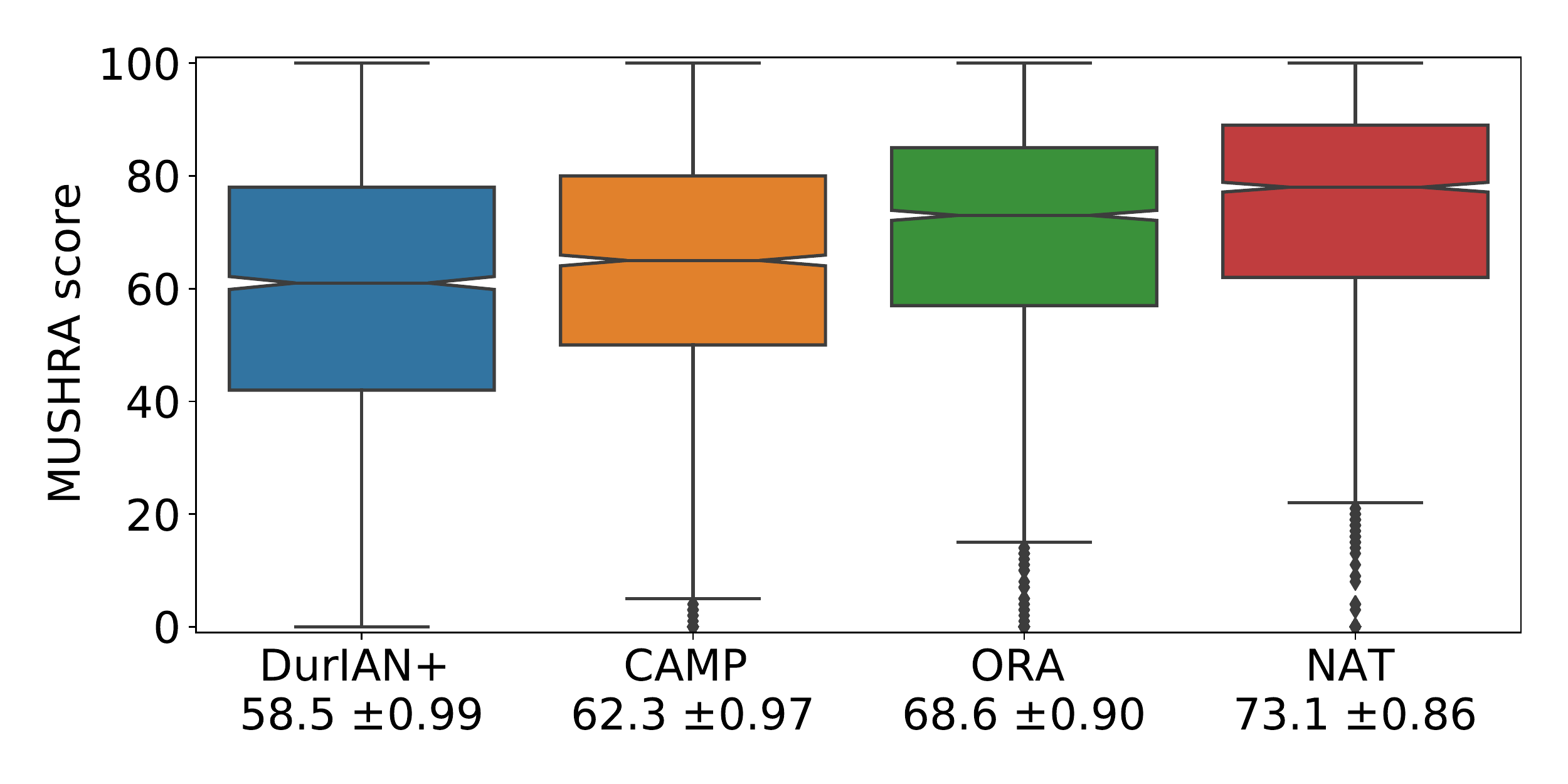}
  \vspace{-20pt}
	\caption{MUSHRA results with \ptm. Mean rating and 95\% confidence intervals are reported below system names. Question asked: \emph{``Rate the systems based on your preference''}.}
  \label{fig:MUSHRA}
  \vspace{-10pt}
\end{figure}

\subsubsection{Prosody prediction}
\label{sec:evaluations-CAMP}
\vspace{-5pt}

Finally, we evaluate \ptm\ against natural speech. We use \autoregressive\ as a strong baseline following the result in Section~\ref{sec:evaluations-AR}. This makes the presence of a duration model a control variable, allowing us to attribute any improvements to the two-stage structure. We perform a MUSHRA-like test \cite{MUSHRA:2014} with 4 systems: \autoregressive, \ptm, \oracle, and \speech.

Since we provide isolated sentences to listeners, there is no concept of correct prosody \cite{rob-SSW19:2019}, as such we do not provide a visible reference and do not require listeners to rate two systems as 0 and 100. 25 listeners completed this test.

We present the results in Figure~\ref{fig:MUSHRA}, along with 95\% confidence intervals, computed using percentile bootstrap. Confidences reported below are computed in the same way. We perform a two-sided Wilcoxon signed-rank test on all 6 pairs of systems, and correct with Holm-Bonferroni. We find all systems are significantly different from each other ($p \ll 10^{-10}$).

From this test we can conclude that: (1) \ptm\ reduces the gap between natural speech and \autoregressive\ by a relative $26\% \pm 7\%$, a significant improvement in prosody quality attributed to our two-stage approach. (2) Given the best-case prediction of the prosody representation (i.e.\ \oracle), our model could close this same gap by a relative $69\% \pm 6\%$. This is below 100\% suggesting, as expected, that our representation is lossy -- likely due to the trade-off between disentanglement and descriptive power. (3) Using BERT as a context encoder, \ptm\ gets $38\%\pm10\%$ of the way to achieving this top-line performance (i.e. the gap between \autoregressive\ and \oracle).

As discussed in Section~\ref{sec:evaluations-AR}, the use of additional context information is not enough to improve prosody modelling. We attribute the performance of \ptm\ to the prosodically relevant loss. The prosody predictor's loss doesn't focus on lower-level acoustics, but directly on prosody. We chose to use a learnt prosody representation, enabling our model to determine which information is important. Explicit prosody features, such as F\textsubscript{0}, could be used instead. However, these may limit the range of prosody, and are susceptible to errors.

The results show a significant improvement in prosody. They also suggest we can improve further within this two-stage paradigm: through improved representation learning; or improved prosody prediction -- using more context features or increased context width (e.g.\ neighbouring sentences).

Our experiments used single-speaker data. Learning a multi-speaker representation would likely improve disentanglement. An investigation of a multi-speaker prosody predictor would be interesting, as this may provide insights into the relationship between different speakers' prosodic patterns.

\vspace{-2pt}
\section{Conclusion}
\label{sec:conclusion}
\vspace{-8pt}

We introduced \ptm, a two-stage approach for prosody modelling. In \emph{stage-1}, we learn a prosodic representation using an encoder-decoder model with a new word-level reference encoder. In \emph{stage-2}, we train a prosody predictor which uses prosodically-relevant context to predict the representations learnt in \emph{stage-1}. We stress the importance for using additional context information when predicting prosody.

An intermediate result showed that replacing attention with a jointly-trained duration model improves significantly over the state-of-the-art, both in terms of stability and prosody. Finally, we show that our proposed approach, \ptm\ -- using BERT to guide prosody prediction -- improves further upon this intermediate result, closing the gap between our stronger baseline and natural speech by 26\%.

\clearpage
{\small
	\bibliographystyle{IEEEbib}
	\bibliography{references}
}

\end{document}